# Considerations Concerning the Overall Unification

Shantilal G. Goradia

Gravity Research Institute, Inc.

Mishawaka, IN USA

**Abstract:** We use entropy to link fine-structure constant and cosmological constant. We also link nuclear force and gravity. We step on the fundamentals of consciousness for this new millennium with a scientific approach. Statistical and quantum mechanical probabilistic approach to explain gravity has a potential to extend physics to other branches of science. We meet the minds of giants such as Newton, Einstein, and Gamow. Implicitly we show gravity as the cumulative effect of all other forces, dominated by the strong force.

## 1.  Introduction

There are innumerable articles on consciousness. We are taking a unique approach. We first explain the mystery of physics by addressing the fundamental constants of nature. Having addressed the constants gives us the clue to consciousness.   Gravity describes the large scale structure of the universe, but fails to describe the microscopic world where observable universe plays its tricks. We propose consciousness is not an exception.  We question any hypothetical existence of any interactions except those of subatomic particles.

## 2.  Inverse Square Law in Natural Units

Newton delayed announcing the law of gravitation. He explained why in Book III, Proposition 8 of his Principia [1]. He was in doubt about the duplicate reciprocal (inverse square law) [2].  A new concept is seen on the horizon. It postulates that the gravitational force between two interacting elementary particles does not change as a function of the classical continuous distance separating them. What changes is the probability of an interaction between elementary particles. What is perceived a continuous gravitational force is a result of probabilities of interactions between a large statistical aggregate of elementary particles that constitute massive bodies such as an apple. Classical distances are manmade units. The use of natural units of Planck lengths replaces the inverse square propagation implicit in Newtonian gravity with inverse square law of probabilities. Consistent with Newtonian gravity, it will be shown that the force between two interacting nucleons is the strongest, known as strong coupling ($C_s$) when two interacting nucleons are closest. The minimum distance per modern physics is one Planck length. It equals $10^{-35}$ meter. Larger distances are integer multiples of a Planck length. The new concept states that the statistical probability function ($P_i$) that two nucleons interact is inversely proportional to the integer number of Planck lengths separating them, making the average effective gravitational force perceived between two interacting nucleons ($F_p$) equal to the product $C_s$ x $P_i$. Nucleons are the most massive elementary particles in an atom. Therefore, they are used in this example to link quantum physics, and Newtonian gravity in one equation.

The Newtonian expression of force between two nucleons separated by a distance r is

$$F(r) = G\, m_n^{\,2}/r^2, \qquad (1)$$

where $F(r)$ is the force in Newton's, $G$ is Universal constant of gravity, $6.672 \times 10^{-11}$ in Newton (meter)$^2$ / (kilogram)$^2$, $m_n$, the mass of a neutron in kilograms, and $r$ is the separating distance in meters.

Conversion of separation distance from r in meters to L in Planck lengths yields:

$$F(L) = k/L^2, \quad (2)$$

where k (constant) = $10^{70} \, G \, m_n^2$.

For $L$ = one Planck length,

$$F(L=1) = k = C_s \quad (3)$$

Since $1/L^2 = P_i$, we get:

$$F(L) = k/L^2 = k \times P_i = F_p, \quad (4)$$

Equation (4) recovers the Newtonian gravity. Its derivation shows that the strongest force between two adjoining nucleons ($C_s$) results naturally when their separation is minimum possible of one Planck length with the probability of interaction equal to one. Rearranging mathematical formulation yields:

$$(F_p) = C_s \times P_i, \quad (5)$$

where the expression $(F_p) = C_s \times P_i$ suggests that the force between two nucleons does not decrease as a function of the distance separating them. It is the probability of interaction that decreases inversely as a function of their separation.

The synonymy of strong coupling and gravitation is elaborated in [3] and open access journal [4], with [3] taking a relativistic approach. Reference [5] links information with entropy. We link Planck scale statistics with information.

## 3. Quantum Informatics

The lack of a Lagrangian linking particles with spacetime creates an unstable link (OPEN and/or CLOSED) between them, capable of sending messages in terms of time-unique qubits (quantum bits) of information. The postulation of particles having quantum (Planck size) mouths, as shown in Figure 1, Figure 2 and Figure 3, for interaction with the rest of the universe explains their ability to send messages. A steady state condition of the mouth would not generate multiple signals. Associated uncertainty may be nature's tool to generate information, and the information so created could have cloned the baby sheep *Dolly*, as elaborated below. If so, the subject postulation may extend the unification efforts to link physics with biology. Biochemistry research shows that molecules exhibit quantum blinking dots exchanging information in terms of statistics. Biologists join the dots [6]. Qubits of information can describe the underlying phenomenon biology is searching for. They may shed light on cosmology as well.

What is the encryption in the blinks? Particles must know, however we do not. Particles must be using varying amounts of energy as allowed by the uncertainty principle to generate information. A cell from bone marrow will, when placed in the liver of a mouse, make liver protein. It must be informed to do so. The qubits of information would be meaningless to human observers ignorant of the encryptions. If we can decode some encryptions, we may be able to use this knowledge to research more effective drugs. Particles aware of the encryption would be able to transmit information. Einstein is reported to believe that "a particle should know where and what it is, […] even if we do not, and it should certainly not receive signals more quickly than the speed of light [7]" Reference number [8] points out noteworthy views of other great minds.

If a tiny cell of an oriental male, whether fertilized in a black woman or a white woman, can clone the entire donor oriental man; it becomes obvious that the trillions of the physical particles of the donor cell must carry its entire genetic information. Whether this biological phenomenon is consistent with any physical theory or not, it is consistent with Einstein's view in [7] about the ability of the particles to know beyond our ability to detect exactly what they know. The known attributes available to an elementary physical particle are its position and velocity which are indeterminate according to the uncertainty principle. Uncertainty principle can be implicated as declaring that we never know what information the particles are exchanging, taking for granted that the particles are exchanging information that enables them collectively to carry out cloning. Einstein did not have the privilege to site cloning, since it was not discovered during his life span. Uncertain positions and velocities of our lips make an understandable lip language to some deaf people who know the spoken language, but not to those creatures who do not know the spoken language. Therefore, those creatures, if they could, would consider the lip movements as strictly probabilistic, supporting one point, while there is no another accepted point, that the uncertainty of the uncertainty principle is a means of communication for the particles expressing coded information which we do not understand.

Qubit is a term used by computer scientist namely Seth Lloyd of MIT [9], supporting an older basic idea with more details. It is proposed hear that particles' indistinguishable positions and velocities fluctuating as mush as every Planck time can be drafted to be equivalent of qubits to enable them to superimpose their OPEN and CLOSED states simply illustrated in the following figure as binary expression for an easier grasp for those who cannot otherwise grasp the computational universe endorsed by Seth Lloyd. Figure 1 gives the pictorial presentation of the idea showing a maximum of four continuous OPEN or CLOSED states.

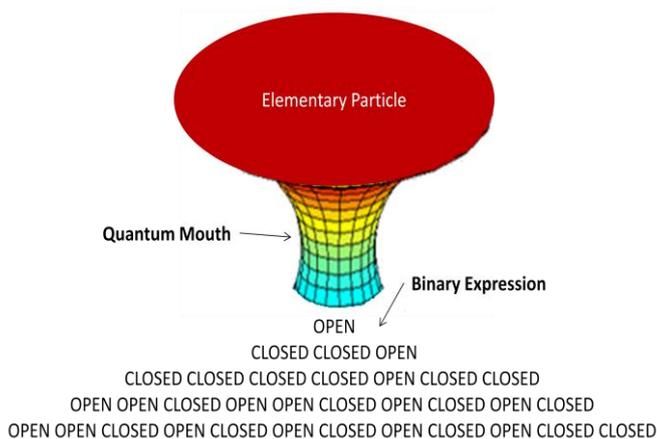

**Figure 1. Mental picture of a particle communicating with the universe via postulated quantum mouth**

Physics uses the universal constant of gravity, G, to determine Planck length. Nature knows the Planck length, a variable or not. Therefore, nature needs no G, making G the net effective resultant force created by all the partial ones by subatomic particles. G is only man's tool to probe the nature, helping him to overcome his otherwise ignorance of Planck length. Mediating particle of G, graviton, has never been observed, we predict: never will be. So long as the efforts are funded, one may keep on trying. Figure 2 depicts a Newtonian equivalent of Figure 1 with gravitons as hypothetical.

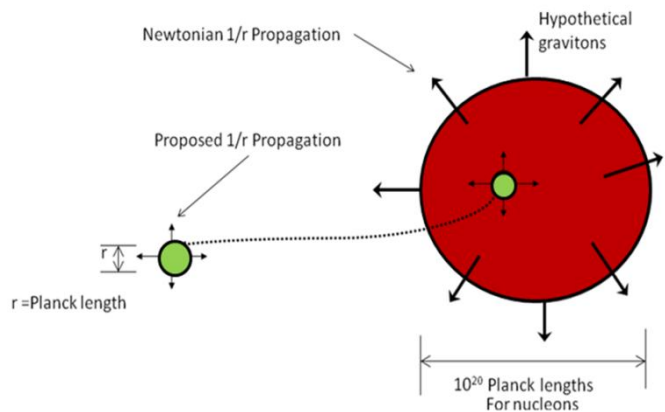

**Figure 2. Mental picture of a quantum mouth radiating hypothetical gravitons**

## 4. Fine-structure Constant

The maximum number of microstates ($W$) the two particles could have created would be $10^{60}$, the number of Planck times per Hubble time $H_t$. Substitution of $W$ in Boltzmann expression, yields the entropy in the Boltzmann expression (on his tomb) $S = k \, Log_e \, W$, to 138 for k =1. Whether the system is binary or ternary, the maximum number of microstate per Hubble time is fixed, as a Planck time cannot be subdivided in 2, 3 or more subdivisions, giving the value of $n$ =1 for substitution in the equation $\Omega = N!/n!(N-n)!$ [10], yielding $\Omega$ (our notation $W$) = $10^{60}$ for $N = 10^{60}$.

There are conjectural and other reasons to believe that the closeness of 138 to the reciprocal (137) of fine structure constant $\alpha$ (1/137) links them together. The discrepancy between 138 and 137 is largely eliminated by a potential correction by physicist Shang-Keng Ma who replaced Boltzmann's law by his own law that divides W by 2 [11]. Future may find other causes, as we do here in section 9, added this year. If Boltzmann did not foresee such an application of his equation, he deserves more credit now than he would have realized while he was alive. Unfortunately, Boltzmann did not live long enough to see his equation vindicated. The fact that Boltzmann equation is a postulate is not a negative, given that all conventional physical theories are based on postulates. Expression of entropy without manmade units is a plus. The use of natural logarithm and natural units reflect truer description of nature than otherwise. Einstein's independent development of statistical mechanics led him to admire Boltzmann statistics [12].

Einstein considered thermodynamics as fundamental [13]. The universe must have some hidden candidate that reflects decrease in entropy without which we would not be here. The likely hidden candidate is the fine-structure constant.

## 5. Cosmological Constant

Gamow saw a connection between cosmology and fine-structure constant [14]. Cosmological constant ($\lambda$) introduced by Einstein in 1917 is a function of the radius of the universe ($R$) [15]. He introduced the equation $\lambda = 1/R^2$. The age of the universe, $10^{60}$ Planck times, translates in the size of $10^{60}$ Planck lengths as light travels one Planck length per Planck time, yielding the following equation.

$$\text{Cosmological Constant } \lambda = 1/(10^{60})^2 \quad (6)$$

The size of the universe is common to the fine-structure constant as derived above and the cosmological constant.

$$\text{Substituting one in the other gives } 1/\alpha = \ln \sqrt{\lambda} \quad (7)$$

The interpretation that entropy is non-decreasing and the fact that the particles used to test the fine-structure constant cannot be older than those used to test the age of the universe justifies the replacement of equality sign by the equal to or greater than signal, giving

$$\alpha \geq 1/\ln \sqrt{\lambda} \quad (8)$$

In his 1919 paper translated in English as now allowed by the present day copyright laws and republished this millennium [16], contrary to the popular belief, Einstein retracted the cosmological constant questioning his own earlier (1917) proposal and Newton's 1686 theory of gravitation. The validity of his 1919 paper is further enhanced by (1) unification of nuclear force and gravitation in section 2, and (2) resolution of Newton's own doubt on the complete validity of his inverse square law. Additional explanations of unexplained observations may come out after publication of this article. It is noteworthy that these true scientists were so strongly committed to science that they did not hesitate to question the part of their own theories that they saw questionable. Einstein's openness made him indiscriminately describe an African high school student as a future physicist to the dislike of the student's teacher.

Alternative cosmology group comes up with several articles every month questioning the status quo. There are a lot of articles by other doctorates, including mechanical engineers who score the highest in GRE scores, questioning the status quo. During conference encounters, their authors say that such articles are not published as they are not consistent with either quantum physics or general relativity or the status quo. How can any article take a lead way to unify if it is completely consistent with a theory that denies unification?

In our draft efforts to juggle through incremental increases in the surface and volume of the universe every Planck time with a focus on the inverse of the cosmological constant introduced by Einstein, we ran into a mathematical series. Some famous relativists have given up attempts to modify general relativity. Whether or not, this mathematical series is any help there, we think it is worth publishing anyway. Here it is: The sum of the cubes of the number series $1^3 + 2^3 + 3^3 \ldots + n^3$ equals the square of their sum $(1 + 2 + 3 \ldots n)^2$.

## 6. Quantum Tunneling

An alpha particle can escape out of its potential well, because there is a probability that the effective quantum mouth (point of collective action of the quantum mouths of its nucleons) lies outside the central portion of the nucleus, on the downward slope as shown in Figure 3. Such simple solutions do not spell incorrect solutions. Many times, the solutions to most complex problems lie in solutions with unthinkable simplicity.

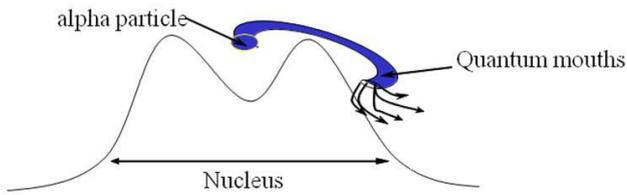

**Figure 3. Postulated mouths on the right lifting the alpha particle**

History has taught us over and over again that those capable of solving complex problems avoid looking at simple solutions. Here is an example for those who know related history. A creative mind could have found a simple solution to the unexplained nature of power fluctuations by pointing out the leaking PORV which caused the biggest accident in USA at Three Miles Island nuclear station that wiped out the nuclear industry in USA at a cost of several hundred billion dollars to the US economy, if not a trillion. Without adding personalized details to this example obtained from personal interview few months before the accident, one point can be made about the importance of simplified solutions. Physicists engaged in the unification of gravity need to look at simplified solutions and encourage their publications first at least in proceedings. Someone keeping on paying for something else does not provide a satisfying scientific answer.

## 7. Fine-structure Constant and Consciousness

Linking the fine-structure constant with natural logarithm fills gaps in other places in science, providing additional conjectural support to the proposition. "The use of natural logarithm has a special significance to the investigation of consciousness. The number *e*, the base of natural logarithm is usually defined as the limit of the expression $(1 + 1/n)^n$ as *n* approaches infinity. One can obtain an expansion of this expression in the following relation.

$$e = 1 + \frac{1}{1} + \frac{1}{1 \cdot 2} + \frac{1}{1 \cdot 2 \cdot 3} + \cdots + \frac{1}{1 \cdot 2 \cdot 3 \cdot n} \quad (9)$$

It can be shown by elementary methods that *e* is irrational; that is, it cannot be represented as the quotient of two integers. Furthermore, *e* is transcendental; it does not satisfy any algebraic equation with integer coefficients. The transcendentality of *e* was proved mathematically by the French mathematician C. Hermite in 1873; the proof constitutes an important milestone in the history of mathematics. The existence of *e* is therefore related to a spiritual or nonphysical realm (soul), incapable of production by algebraic manipulations as promulgated by an international organization founded by guru Maharishi Mahesh Yogi [17]".

Is it just a coincidence that a logarithmic formula extends its utility to biology? Future research will tell. For now, it is worthwhile to note: "Inasmuch as the mathematical relationship be derivable from psychoneural and neurophysical equations, Fechner's panpsychist interpretation of his psychophysical equation assumed that the underlying psychoneural equation was a logarithmic equation yielding the negative sensations and that the neurophysical equation was linear [18]". Do particles unite to make memory that is part of living bodies?

## 8. The Cause of Memory

"Precisely how we store trillions of bits of information – more than any computer can store – remains one of the mysteries of the mind. Scientists are equally baffled by how we lose that information. They do know this: New memories are encoded in the brain's hippocampus. Scientists suspect memories to go from there to the frontal lobes [19]". How does memory move? There are no memory particles. What causes the movement in our opinion is that the changed ability of the quantum particles to interact remotely at distances that are changed due to age related structural changes and other issues. Without the probabilistic propagation of Newtonian gravity as modified here, we fine it difficult to show the diminishing probability of particle interactions as a consequence of the structural changes in the brain.

Alzheimer's known link to structural changes in the brain by cerebral arteriosclerosis, physical blow, plaque formation, inflammation, infection or other age related causes would make gravitational changes too difficult to check experimentally, but degradation of neurons can be better supported by probabilistic aspect of gravity than by Newtonian gravity, further supported by the lack of observation of any graviton. The idea of particles interacting remotely with a quantum mouth is now found to be somewhat analogous to the one in string theory. "Most recent models including Quantum Superstrings on the contrary deal with extended and not point particles….[20]" When we investigate the shape of a particle from the closest vicinity, we get the most probable shape of the particle, not its true permanent shape. The shape of a particle is affected by the particles of the body that test its shape.

## 9. Imaginary Dimensions of String Theory

String theory tries to overcome the inability of quantum physics. Despite of all of its successes, it cannot come up with a proof of a realistic existence of multiple dimensions, because there are only three observable dimensions and time.

Regardless, the number of imaginary dimensions, 6 or 10, used by two different string theories may support the discrepancy between 138 and the reciprocal of the fine structure constant (137). The continuous OPEN or CLOSED position of the quantum mouth for few (6 or 10) Planck times may not make more than one bit of information by nature. Possibility cannot be excluded that few of there continuous OPEN or CLOSED positions make a super bit (BIT, if I may) of nature. Then the number of BITS would be lightly less than $10^{60}$ bits, bringing 138 closer to 137. String theory fails to give exact results and different theories uses different number of imaginary dimensions.

A BIT may not have an exact number of bits. Ever changing ratio of BIT/bit can bring about evolution without destabilizing the constancy of the fine structure constant. Einstein was made aware of the constancy of the fine structure constant in 1916 by his friend Arnold Sommerfeld [21], before he introduced the cosmological constant in 1917, whether or not his decision was influenced by the finding of the constancy of the cosmological constant.

## 10. Meditation

Meditation is no longer a subject of religion alone as is now well proven by medical science. How did the ancient Vedic Philosophy (not any religion) describe the particles as *anu-atma* [particle soul], further stating them as knowledgeable (*vipascin*) and also quantify their size (ten thousands of the tip of hair [atom] equal to the size of a nucleon, a subatomic particle? They did not know subatomic physics. "Per ancient Vedas (*Katha* Upanishad 1.2.20), a particle has a particle-soul (*anu-atma*), and it is also connected to the Supreme, which is omnipresent (*param-atma*). I see some qualitative consistency between this Vedic doctrine and my proposal based on a potentially justifiable speculation that anu-atma and param-atma are analogous to particle and normal space-time respectively, connected by some quantum entity [22]".

The reason why we cannot decide the position and velocity of a particle accurately is that the particle must know what we do not as somehow claimed by Einstein [7]. In the last paragraph of Principia, Newton refers to "Spirit of the particles" and wonders about their biological functions [1]. With due concentration, the particles within us must be revealing what we do not know. If not, how else can we support wonders of meditation? This approach, unless challenged, leads to a fundamental link between science and spirituality. How can we rule out that the cumulative effect of the unrecognizable consciousness of the particles is the recognizable consciousness of living bodies? If we justifiably look at physics as an advanced form of philosophy, and further extend the result to compare with diversity of views over the globe, sometimes, in some aspects and somewhat we suspect: "Religion may be a result of meditation [23]".

## 11. Conclusion

One of the implicit remarks is that particles have a quantum mouth, and a body they can stretch, and get entangled. There is no compelling reason to stick to the 300 year old notion that gravity is a fundamental interaction even if so suggested by some theories. We have to ask: Do such theories unify?

## 12. References

[1] Newton, I.; *The Principia*, Prometheus Books: Amherst, Mass, USA, 1995, pp. 334.

[2] Cajori, F.; Newton's twenty years delay in announcing the law of gravity, in *Sir Isaac Newton 1727-1927*, Williams and Wilkins Company: Baltimore, USA, 1928, pp. 127-188.

[3] Shrivastava S. K. *Aspects of Gravitational Interactions* Nova Science Publishers: Hauppauge, USA, 1998, p. 91.

[4] Goradia S. *Why is Gravity so Weak?* Journal of Nuclear Radiations and Physics, 1, pages 107-117, 2006. See also: http://www.arXiv.org/pdf/physics/0210040v2.

[5] Duncan T and Semura J. *The Deep Physics Behind the Second Law: Information and Energy As Independent Forms of Bookkeeping*, Entropy, 6, pages 21-29, 2004.

[6] Klarreich E., *Biologists join the dots*, Nature, 413, pages 450-452, 2001.

[7] Minkel, J.R., *The Gedanken Experimenter*, Scientific American, 297, pages 94-96, 2007.

[8] Goradia, S., *What is Fine-structure Constant?;* http://www.arXiv.org/pdf/physics/0210040v3.

[9] Lloyd Seth, *Programming the Universe*, Division of Random House, Inc., New York, USA, 2006, pp. 109-118.

[10] Glazer M. and Wark J., *Statistical Mechanic: A Survival Guide*, Oxford University Press Inc. New York, USA, 2001, pp. 5-8.

[11] Berlinski D.; *The Advent of the Algorithm.* Harcourt Inc.; Orlando, Florida, USA, 2001, p. 231.

[12] Charles S., *Decoding the Universe: How the New Science of Information is Explaining Everything in the Cosmos, from Our Brains to Black Holes*, Penguin Group Inc., New York, USA, 2006, p.121.

[13] Einstein, A. *Considerations Concerning the Fundamentals of Theoretical Physics*. Science, 91, 2369, pages 487-492, 1940.

[14] Gamow G., *Electricity, Gravity and Cosmology*, Phys. Rev. Lett., 19, pages 759-761, 1961.

[15] Einstein, A.; *Cosmological Considerations on the General Theory of Relativity*. In: *The Principle of Relativity*, Dover Publications: Mineola, USA, 1952, pp. 177-188.

[16] Einstein, A.; *Do Gravitational Fields Play an Essential Part in the Structure of the Elementary Particles of Matter?* In: *The Principle of Relativity*, Dover Publications: Mineola, USA, 1952, pp. 189-198.

[17] Lowan, A.; Bochner, S.; *e*(mathematics) In: *McGraw-Hill Encyclopedia of Science & Technology 5 (Cot-Eat),* McGraw-Hill: New York, USA, 2007, p. 749.

[18] Baars, B.J., Consciousness In: *McGraw-Hill Encyclopedia of Science & Technology 4 (Che-Cos),* McGraw-Hill: New York, USA, 2007, pp. 549-552.

[19] Elizabeth Leland, MeClatchy Newspapers In: South Bend Tribune, USA 2010, p. D6.

[20] Sidharth, B. G., *Alternative Routes to Gravitation.* In: *Frontiers of Fundamental and Computational Physics* – AIP Conference Proceedings, New York, USA, 2010, pp. 64-74.

[21] Fritzsch H., *The fundamental Constants, A mystery of physics*, World Scientific Publishing Co. Singapore. 2009, p. 9

[22] Goradia, S. G., *My Focus on the Quantum Source of Gravity, Frontiers of Fundamental Physics*, Springer, Dordrecht, Netherlands, 2006, pp. 37-44.

[23] Goradia, S. G., *Vedic Science and Quantum Physics, Volume 4, Spirituality, World Religions After September 11,* PRAEGER, Westport, USA, 2009, pp. 147-152.